\documentclass{article} 
        \newcommand{\tr}{{\rm tr}}   
           
	\newcommand{\be}{\begin{equation}}   
        \newcommand{\ee}{\end{equation}}   
        \newcommand{\ba}{\begin{eqnarray}}   
        \newcommand{\ea}{\end{eqnarray}}   
        \newcommand{\ban}{\begin{eqnarray*}}   
        \newcommand{\ean}{\end{eqnarray*}}

\usepackage{latexsym,amsfonts,amsthm,amssymb,exscale,enumerate}
\usepackage[centertags,sumlimits,intlimits,namelimits,reqno]{amsmath}

\usepackage{xcolor}
\usepackage{framed,color}

\definecolor{shadecolor}{rgb}{1,0.8,0.3}
\definecolor{myurlcolor}{rgb}{0.5,0,0}
\definecolor{mycitecolor}{rgb}{0,0,0.8}
\definecolor{myrefcolor}{rgb}{0,0,0.8}
\definecolor{hyperrefcolor}{rgb}{0.7,0,0}

\usepackage[bookmarks=false]{hyperref}
\hypersetup{
	colorlinks,
	linkcolor={myrefcolor},
	citecolor={mycitecolor},
	urlcolor={myurlcolor}
}

\begin{document}   
 
	\begin{center}   
	{\bf R\'enyi Entropy and Free Energy \\$ \; $ \\}   
	{\em John\ C.\ Baez\\}   
	\vspace{0.3cm}   
	{\small Centre for Quantum Technologies  \\
        National University of Singapore \\
        Singapore 117543  \\  
        \vspace{0.3cm}
        and \\ 
        \vspace{0.3cm}
        Department of Mathematics \\   
	University of California\\   
        Riverside CA 92521\\}   
        \vspace{0.3cm}
        {\small email:  baez@math.ucr.edu\\} 
	\vspace{0.3cm}   
	{\small \today}
	\vspace{0.3cm}   
	\end{center}   

\begin{abstract}
\noindent The R\'enyi entropy is a generalization of the usual concept
of entropy which depends on a parameter $q$.  In fact, R\'enyi 
entropy is closely related to free energy.  Suppose
we start with a system in thermal equilibrium and then 
suddenly divide the temperature by $q$.  Then the maximum amount of work the
system can do as it moves to equilibrium at the new temperature,
divided by the change in temperature, equals the system's R\'enyi 
entropy in its original state.  This result applies to both classical and 
quantum systems.  Mathematically, we can express this result as follows:
the R\'enyi entropy of a system in thermal equilibrium is minus the 
`$q^{-1}$-derivative' of its free energy with respect to temperature.  
This shows that R\'enyi entropy is a $q$-deformation of the usual 
concept of entropy.  
\end{abstract}

\section{Introduction}

In 1960, R\'enyi \cite{Renyi} defined a generalization of Shannon entropy 
which depends on a parameter.  If $p$ is a probability distribution on 
a finite set, its R\'enyi entropy of order $q$ is defined to be
\be
\label{renyi}
S_q= \frac{1}{1 - q} \ln \displaystyle{\sum_i p_i^q} 
\ee
where $0 < q < \infty$.   Of course we need $q \ne 1$ to avoid
dividing by zero, but L'H\^opital's rule shows that 
the R\'enyi entropy approaches the Shannon entropy as $q$ approaches
1:
\[
       \lim_{q \to 1} S_q = -\sum_{i} p_i \ln p_i .
\]
Thus, it is customary to define $S_1$ to be the Shannon entropy.  

While Shannon entropy has a deep relation to thermodynamics, 
the R\'enyi entropy has not been completely integrated into this 
subject---at least, not in a well-recognized way.  
While many researchers have tried to {\it modify} statistical mechanics
by changing the usual formula for entropy, so far
the most convincing uses of R\'enyi entropy in physics seem 
to involve the limiting cases $S_0 = \lim_{q \to 0} S_q$ and 
$S_\infty = \lim_{q \to +\infty} S_q$.  These are 
known as the `max-entropy' and `min-entropy', respectively, 
since $S_q$ is a decreasing function of $q$.
They show up in studies on the work value of information \cite{DRRV} 
and the thermodynamic meaning of negative entropy \cite{ADRRV}.  
For other interpretations of R\'enyi entropy see Harrem\"oes \cite{Harremoes},
K\"onig {\it et al.} \cite{KRS}, and Uffink \cite{Uffink}.

In fact, it is not necessary to modify statistical mechanics to 
find a natural role for R\'enyi entropy in physics.  R\'enyi 
entropy is closely related to the familiar concept of 
{\it free energy}, with the parameter $q$ appearing as
a ratio of temperatures.  

The trick is to think of the probability distribution as a Gibbs
state: that is, the state of thermal equilibrium for some 
Hamiltonian at some chosen temperature, say $T_0$.  Suppose that 
all the probabilities $p_i$ are nonzero.  Then working in
units where Boltzmann's constant equals 1, we can write
\[      p_i  = e^{-E_i/T_0} \]
for some nonnegative real numbers $E_i$.  If we think of these
numbers as the energies of microstates of some physical system, 
the Gibbs state of this system at temperature $T$ is the probability 
distribution 
\[
       \frac{e^{-E_i/T}}{Z(T)} 
\]
where $Z$ is the partition function:
\[
      Z(T) = \sum_{i \in X} e^{-E_i/T}  
\]
Since $Z(T_0) =1$, the Gibbs state reduces to our 
original probability distribution $p$ at this temperature.  

Starting from these assumptions, the free energy 
\[
F(T) = - T \ln Z(T)
\]
is related to the R\'enyi entropy as follows:
\be
       F(T) = - (T - T_0) S_{\, T_0/T}
\label{preliminary_relation}
\ee
The proof is an easy calculation:
\[
S_{\,T_0/T} = \frac{1}{1 - T_0/T} \ln \sum_{i} p_i^{T_0/T} = 
\frac{T}{T - T_0} \ln \sum_{i} e^{-E_i/T} = -\frac{F(T)}{T - T_0} .\]
This works for $T \ne T_0$, but we can use L'H\^opital's rule to show
that in the limit $T \to T_0$ both sides converge to the Shannon entropy
$S_1$ of the original probability distribution $p$.  

After the author noticed this result in the special case $T_0 = 1$
\cite{Azimuth}, Stacey commented that this case was already
mentioned in Beck and Schl\"ogl's 1995 text on the thermodynamics
of chaotic systems \cite{BS}. However, most people using R\'enyi entropy 
were unaware of its connection to free energy, perhaps because they work 
on statistical inference rather than physics \cite{EX}.   Thus, the author put
a version of this note on the arXiv in 2011 \cite{Baez}. It has subsequently been 
cited 78 times, which suggests that it was indeed useful.  We have therefore 
decided to publish it.

Shortly after the first draft of this note was released,
Polettini gave a nice physical 
intepretation of R\'enyi entropy \cite{Polettini2}.
Downes then made a further generalization \cite{Downes}.  
The above argument concerns a system with Gibbs state $p_i = 
\exp(-E_i/T_0)$ at a chosen temperature $T_0$.  Such a system
automatically has zero free energy at this chosen temperature.
Downes generalized the relation between R\'enyi entropy and free energy 
to systems whose free energy is not constrained this way.  
The physical interpretation of R\'enyi entropy can be
extended to these more general systems, and we describe this 
interpretation in what follows.  We also explain how the R\'enyi entropy
is a `$q$-deformation' of the ordinary notion of entropy.  This
complements the work of Abe on another generalization of entropy,
the Tsallis entropy \cite{Abe}.

In what follows, we work in a quantum rather than classical
context, using a density matrix instead of a probability distribution.
However, we can diagonalize any density matrix, and then its
diagonal entries define a probability distribution.  Thus, all
our results apply to classical as well as quantum systems.
The quantum generalization of Shannon entropy is, of course, well-known:
it is the von Neumann entropy.  The quantum generalization
of R\'enyi entropy is also already known \cite{vDH}.

\section{R\'enyi Entropy as a $q$-Derivative of Free Energy}

Let $H$ be a self-adjoint complex matrix.  Thinking 
of $H$ as the Hamiltonian of a quantum system, and no longer assuming that 
Boltzmann's constant $k$ equals 1, we may define the Gibbs state of
this system at temperature $T$ to be the density matrix
\be 
\label{gibbs}
\rho_T = \frac{1}{Z(T)} e^{-H/kT} 
\ee
where the partition function
\be
\label{partition}
Z(T) = \tr(e^{-E/kT})
\ee
ensures that $\tr(\rho_T) = 1$.   The Helmholtz free energy at temperature
$T$ is defined by
\be  F(T) = -kT \ln Z(T)  . \ee

On the other hand, for any density matrix $\rho$, the quantum 
generalization of R\'enyi entropy is defined by
\be
S_q(\rho) = k \; \frac{\ln \tr(\rho^q)}{1 - q}  
\label{quantum_renyi}
\ee
since this formula reduces to the usual definition of R\'enyi
entropy, Equation (\ref{renyi}), when the probabilities 
$p_i$ are the eigenvalues of $\rho$ and we set $k = 1$.
This formula makes sense when $0 < q < \infty$ and
$q \ne 1$, but we can define the quantum R\'enyi entropy as a limit
in the special cases $q = 0, 1 , +\infty$.  For $q = 1$
this gives the usual von Neumann entropy:
\be
 S_1(\rho) := 
\lim_{q \to 1}  S_q(\rho) = -k \, \tr(\rho\ln \rho) .
\label{von_Neumann}
\ee

Returning to our system with Gibbs state $\rho_T$ at temperature
$T$, let us write $S_q(T)$ for $S_q(\rho_T)$.   Computing this
R\'enyi entropy at some temperature $T_0$, we find:
\ban S_q(T_0) &=& k\,  \frac{\ln \tr(\rho_{T_0}^q)}{1-q} \\
&=&  \frac{k}{1 - q} \,  \ln \tr\left(\frac{e^{-q H/T_0}}{Z(T_0)^q} \right) \\
&=& \frac{k}{1-q} \big(Z(T_0/q) - q Z(T_0)  \big) 
\ean 
If we define a new temperature $T$ with 
\be        q = T_0/T , \ee
we obtain:
\ban  S_q(T_0) 
&=& k\, \frac{\ln Z(T) \; - \; q \ln Z(T_0)}{1 - q} \\   \\
&=& k\, \frac{T \ln Z(T) \; - \; T_0 \ln Z(T_0)}{T - T_0} \\   \\
\ean
or in short:
\be
S_{T_0/T}(T_0) = -\; \frac{F(T) - F(T_0)}{T - T_0} .
\label{fundamental_relation}
\ee

This equation is the clearest way to
express the relation between R\'enyi entropy and free energy. 
In the special case where the free energy vanishes at temperature
$T_0$, it reduces to Equation (\ref{preliminary_relation}).
In the limit $T \to T_0$, it reduces to
\be
S_1(T_0) = - \left.\frac{d F(T)}{d T} \right|_{T = T_0} .
\ee
Of course, it is already well-known that the von Neumann entropy
is the derivative of $-F$ with respect to temperature.  
What we see now is that the R\'enyi entropy is the difference quotient
approximating this derivative.  Instead of the slope of the tangent
line, it is the slope of the secant line.

In fact, we can say a bit more: the R\'enyi entropy is the
the `$q^{-1}$-derivative' of the negative free energy. 
For $q \ne 1$, the $q$-derivative of a function $f$ is defined by
\[   \left(\frac{df}{dx}\right)_q = \frac{f(qx) - f(x)}{qx - x}  .\]
This reduces to the ordinary derivative in the limit $q \to 1$.  
The $q^{-1}$-derivative is defined the same way but with $q^{-1}$
replacing $q$.  
Equation (\ref{fundamental_relation}) can be rewritten more tersely 
using this concept:
\be
S_q =
- \left(\frac{dF}{dT}\right)_{q^{-1}} 
\label{fundamental_relation_2}
\ee
Here we have made a change of variables, writing $T$ for the variable
called $T_0$ in Equation (\ref{fundamental_relation}).  

The concept of $q$-derivative shows up in mathematics 
whenever we `$q$-deform' familiar structures, obtaining new ones
such as quantum groups.  For an introduction, see the text by 
Cheung and Kac \cite{CK}.  In some cases $q$-deformation should 
be thought of as quantization, with $q$ playing the role of 
$\exp(\hbar)$.  That is definitely not the case here: 
the parameter $q$ in our formulas is unrelated to Planck's 
constant $\hbar$.  Indeed, Equation (\ref{fundamental_relation_2})
holds in classical as well as quantum mechanics.  

What, then, is the thermodynamic meaning of R\'enyi entropy?
This was pointed out by Polettini \cite{Polettini2}.  
Start with a physical system in thermal equilibrium 
at some temperature.  Then `quench' it, suddenly 
dividing the temperature by $q$.  The maximum amount of work the 
system can do as it moves to thermal equilibrium at the new temperature,
divided by the change in temperature, equals the system's R\'enyi 
entropy of order $q$ in its original state.   Note that this formulation
even accounts for the minus sign in Equation (\ref{fundamental_relation}),
because it speaks of the work the system does, rather than the
work done to it.

\subsubsection*{Acknowledgements}

I thank all the members of the Entropy Club at the Centre for 
Quantum Technologies, and especially Oscar Dahlsten,
for exciting discussions in which these ideas emerged.  
I thank David Corfield and Tom Leinster for useful discussion
of R\'enyi entropy, and Blake Stacey for pointing out references 
relating it to free energy.  I especially thank Matteo Polettini 
and Eric Downes for suggestions that vastly improved this paper.

\end{document}